# Distributed Universally Optimal Strategies for Interference Channels with Partial Message Passing

Vaneet Aggarwal

Department of Electrical Engineering,

Princeton University,

Princeton, NJ 08544.

vaggarwa@princeton.edu

Salman Avestimehr

Department of ECE,

Cornell University,

Ithaca, NY 14853.

avestimehr@ece.cornell.edu

Ashutosh Sabharwal

Department of ECE,

Rice University,

Houston, TX 77005.

ashu@rice.edu

*Abstract*—In distributed wireless networks, nodes often do not know the topology (network size, connectivity and the channel gains) of the network. Thus, they have to compute their transmission and reception parameters in a distributed fashion. In this paper, we consider that each of the transmitter know the channel gains of all the links that are at-most two-hop distant from it and the receiver knows the channel gains of all the links that are three-hop distant from it in a deterministic interference channel. With this limited information, we find a condition on the network connectivity for which there exist a distributed strategy that can be chosen by the users with partial information about the network state, which achieves the same sum capacity as that achievable by the centralized server that knows all the channel gains. Specifically, distributed decisions are sum-rate optimal only if each connected component is in a one-to-many configuration or a fully-connected configuration. In all other cases of network connectivity, the loss can be arbitrarily large.

## I. INTRODUCTION

One of the fundamental challenges in mobile wireless networks is lack of complete network state information with any single node. In fact, the common case is when each node has a partial view of the network, which is different from other nodes in the network. As a result, the nodes have to make distributed decisions based on their own local view of the network. One of the key question then arises is how often do distributed decisions lead to globally optimal decisions.

The study of distributed decisions and their impact on global information-theoretic sum-rate performance was initiated in [1] for two special case of deterministic channels, and then extended to Gaussian version of those topologies in [2]. The authors proposed a protocol abstraction which allows one to narrow down to relevant cases of local view per node. The authors proposed a message passing protocol, in which both transmitters and receivers participate to forward messages regarding network state information to other nodes in the network. The local message-passing allows the information to trickle through the network and the longer the protocol proceeds, the more they can learn about the network state. More precisely, the protocol proceeds in rounds, where each round consists of a message by each transmitter followed by a message in response by each receiver. Half rounds are also allowed, where only transmitters send a message. One of the main results in [2] is that with 1.5 rounds of messaging, the gap between network capacity based on distributed decisions and that based on centralized decisions can be arbitrarily large for a three-user double-Z channel (two Z-channels stacked on each other). Thus, for some channel gains, decisions based on the nodes' local view can lead to highly suboptimal network operation.

In this paper, we consider the general problem of single-hop $K$-user deterministic interference channels [3–7] with arbitrary network connectivity. Our key result is a complete characterization of all topologies which can be universally optimal with 1.5 rounds of messaging. A scheme is considered to be universally optimal if for all channel gains, the distributed decisions lead to sum-rate which is same as sum-capacity with full information. With 1.5 rounds of messaging, a transmitter knows all channel gains which are two hops away from it and a receiver knows all gains which are three hops away from it. So if the network diameter is larger than three, then no node in the network has full information about the network. Thus, while the capacity of general interference channel is still unknown (even with full network state information), we can characterize which topologies can be universally optimal with partial information.

It turns out that only those networks whose connected components are either fully-connected or have one-to-many connectivity can be universally optimal. The result is intuitively satisfying, since in both cases the nodes which need to control their transmissions to balance their own rate and interference to other receivers have *full* information about the network after 1.5 rounds. For the proof of non-existence of a universally optimal strategy, we provide the global topology information as the genie which each of the node can use to make decisions. For achievability, we give a strategy for any local topology knowledge which would be optimal when there exist a universally optimal strategy for the global topology.

The rest of the paper is organized as follows. In Section II, we formulate the problem and give some definitions that will be used throughout the paper. In Section III, we present our main results and Section IV concludes the paper.

## II. PROBLEM FORMULATION

Consider a deterministic interference channel with $K$ transmitters and $K$ receivers, where the inputs at $k^{\text{th}}$ transmitter in time $i$ can be written as $X_k[i] = \begin{bmatrix} X_{k_1}[i] & X_{k_2}[i] & \ldots X_{k_q}[i] \end{bmatrix}^T$, $k = 1, 2, \cdots, K$, such that $X_{k_1}[i]$ and $X_{k_q}[i]$ are the most and the least significant bits, respectively. The received signal of user $j$, $j = 1, 2, \cdots, K$, at time $i$ is denoted by the vector $Y_j[i] = \begin{bmatrix} Y_{j_1}[i] & Y_{j_2}[i] & \ldots & Y_{j_q}[i] \end{bmatrix}^T$. Associated with each transmitter $k$ and receiver $j$ is a non-negative integer $n_{kj}$ that defines the number of bit levels of $\mathbf{X}_k$ observed at receiver $j$. The maximum level supported by any link is $q = \max_{j,k}(n_{jk})$. The network can be represented by a square matrix $H$ whose $(j, k)^{th}$ entry is $n_{jk}$. Note $H$ need not be symmetric.

Specifically, the received signal $Y_j[i]$ is given by

$$Y_j[i] = \sum_{k=1}^{K} \mathbf{S}^{q-n_{kj}} X_k[i] \qquad (1)$$

where $\oplus$ denotes the XOR operation, and $\mathbf{S}^{q-n_{jk}}$ is a $q \times q$ shift matrix with entries $\mathbf{S}_{m,n}$ that are non-zero only for $(m, n) = (q - n_{jk} + n, n), n = 1, 2, \ldots, n_{jk}$.

We next define the notion of *topology*. We assume that that there is a direct link between every transmitter and its intended receiver. On the other hand, if a cross-link between transmitter $i$ and receiver $j$ does not exist, then $H_{ij} \equiv 0$. Thus, a topology $\mathcal{T}$ is a set of weighted graphs defined as

$$\mathcal{T}(I) = \{H : H_{ij} \equiv 0 \text{ if } (i, j) \in I \text{ else } H_{ij} \in \{0, 1, \ldots, q\}\}.$$

Note that the channel gain can be zero but not guaranteed to be if the index pair $(i, j) \notin I$[1].

We assume that none of the channel coefficients in the matrix $H$, or even the size of matrix $H$ is known before the start of the message passing protocol. As a result, none of the nodes are aware of the maximum possible transmission rates and the associated coding schemes to achieve the capacity. The decision taken by the nodes only depend on the information that the nodes possess. We use the message passing protocol as mentioned in [2] for 1.5 rounds which gives each transmitter the knowledge of all the channel gains of the links that are at-most two-hops away from it and each receiver the knowledge of all the channel gains of the links that are at-most three-hops away. We also assume that the message passing algorithm passes the identities of the end nodes of a link with the information of the link channel gain. With this knowledge of local topology (including node identities) and local channel gains, we answer the question what topology admit existence of strategies that the various transmitters can decide based on the local information that would be sum-rate optimal for all the choices of the channel parameters that are not known to the transmitter. More formally, we answer if there exist an universally optimal strategy with limited information the nodes possess which is defined as follows.

**Definition 1** ([2]). *A universally optimal strategy for a network with d or d.5 rounds of message passing is defined as the strategy that each of the transmitter uses based on its local information in a distributed fashion, such that there exist a sequence of codes having rates $R_i$ at the transmitter $i$ such that the error probabilities at the receivers $\lambda_1(n), \cdots \lambda_K(n)$ go to zero as $n$ goes to infinity, satisfying*

$$\sum_i R_i = C_{sum}$$

*for all the choices of channel gains, where $C_{sum}$ is the sum-capacity of the whole network with the full information.*

---

[1]This is modeled since in a fading channel, the existence of link is based on its average channel variance while the link gain is instantaneous channel value. Thus, the link may on an average be good but its instantaneous value may be below the resolution. The channel information can pass through all the edges in $I$.

## III. Existence of Universally Optimal Strategies with 1.5 rounds of Message Passing Protocol

In this section, we find the condition on topologies for which there exist a universally optimal strategy with 1.5 rounds of message passing. To give the condition, we first define the two sets of configuration of topologies called one-to-many configuration or a fully-connected configuration.

**Definition 2.** *A topology of $K$ users is in one-to-many configuration if there are $2K-1$ links in the topology that include one of the transmitters connected to all the receivers while all other transmitters only connected to their own receivers.*

**Definition 3.** *A topology of $K$ users is in fully-connected configuration if there are $K^2$ links in the topology with each transmitter connected to all the receivers.*

The next theorem describes our main result. Suppose that each node only knows the local topology information, local channel gains and the local node identities. We find the topologies for which a universally optimal strategy exist. The outer bound provides a genie aided topology information to all the nodes while the achievable strategy assumes only the local information.

**Theorem 1.** *Suppose that each node knows only the network topology (or the global network connectivity) information provided by 1.5 rounds of message passing protocol. Then, there exist a universally optimal strategy for a $K$-user interference channel with 1.5 rounds of message passing if and only if all the connected components of the topology are in one-to-many configuration or fully-connected configuration.*

*Proof:* We will first prove that the topologies in which there is a connected component that is not in one-to-many configuration or in fully-connected information, universally optimal strategy does not exist. We first note that for $K < 3$, all the topologies have connected components that satisfy the property in the statement of the theorem and thus the result holds trivially.

The theorem has been shown for $K = 3$ in Appendix A. We will now consider $K > 3$. Consider that there exist a connected component with $K > 3$ users which is not in the one-to-many configuration or in the fully-connected configuration. Then, two cases arise:

1) There exist a transmitter (say $T_1$) which has degree $d$ satisfying $1 < d < K$.

2) All the transmitter nodes have degrees 1 or $K$, such that the number of nodes $n$ having degree $K$ satisfy $1 < n < K$.

For the first case, take the nodes $1, \cdots, d$ as the nodes whose receivers are connected to $T_1$. Now, there exist a transmitter-receiver pair among $d+1, \cdots, K$ whose transmitter or receiver is connected to any of the nodes $1, \cdots, d$. Choose any such pair and call it pair $d+1$. The receiver of $d+1$ is not connected to transmitter 1. Now if the receiver of first is connected to the transmitter of $d+1$, then choose the nodes $1, 2, d+1$ and assume that the direct link of all other users is zero and this information is given as a genie to all the nodes. This creates a genie-aided system in which the nodes 1, 2 and $d+1$ have the uncertainties about all the links connecting them and know 2-hops of information among these links only. In this genie-aided system, there does not exist any universally optimal strategy thus proving the claim (since it makes a connected three-user component which is not in the one-to-many configuration or in the fully-connected configuration). If pair $d+1$ is not connected to pair 1, let us say it is connected to pair $2 \leq j \leq d$. Then, choosing nodes 1, $j$, $d+1$ and repeating the same argument as above proves the statement.

For the second case, choose the three nodes as any two nodes in which the transmitter has degree $K$ and one in which the transmitter has degree 1. Repeating the above genie-aided proof for these three nodes proves the theorem.

This completes the proof that there does not exist a universally optimal strategy for a topology that does contain a connected component which is not in the one-to-many configuration or in the fully-connected configuration.

For the achievability, consider the following strategy. Consider the following cases of the local topology information seen by a user.

1) One-to-many topology with $L$ nodes and the current node has degree 1: The transmitter sends at a rate of $n_{ii}$.

2) One-to-many topology with $L$ nodes and the current node has degree $L$: The transmitter sends at the signal levels that do not potentially create interference to all the users that it interferes.

3) Fully connected topology with $L$ nodes: The node uses the node identities to get its ordering in $L$ nodes and uses a pre-decided strategy that will be optimal for fully connected $L$ node topology.

4) Any other local information: The node sends at a rate

0, or in other words remain silent.

First, it is easy to see that this strategy is optimal if all the connected components of the topology are in one-to-many configuration or fully-connected configuration. For fully-connected components, all the nodes know their connected component and thus can do optimal for its component. For the one-to-many components, each of the users whose transmitters have degree 1 send at rate equal to the rate that the direct channel can support and the remaining user knows all the channel gain and adjust its rate correspondingly. Assume that it is one-to-many component of $L$ users with the first transmitter having degree $L$. The above strategy achieves a sum rate of

$$R_{sum} = \sum_{i=2}^{L} n_{ii} + \sum_{i=1}^{n_{11}} \mathbf{1}_{|U_k|=0}, \qquad (2)$$

where $|U_k|$ is the number of users potentially experiencing interference from the $k^{th}$ signal level of first transmitter which is the same as the sum capacity with global channel information in [7].

Further, it is also easy to see that if using this strategy, we remove the links connected to all the users that are not transmitting, the equivalent topology has all the connected components in one-to-many configuration or fully-connected configuration and in both the cases the data can be decoded. Thus, this strategy is achievable for all possible topologies and is optimal when all the connected components of the topology are in one-to-many configuration or fully-connected configuration thus proving the Theorem. ∎

## IV. CONCLUSIONS

We consider a general deterministic interference channel in which the nodes know the channel gains through a message passing protocol. With 1.5 rounds of this protocol, each transmitter learns the channel gains of all the links that are at-most two-hop distant from it and the receivers learn the channel gains of all the links distant at-most three-hops from it. With this limited information, this paper classifies all interference channel topologies based on their ability to support distributed strategies which are universally optimal. We also note that the genie-aided information regarding global network connectivity do not aid in making a topology able to support a universally optimal strategy.

The problem of defining the exact graph-theoretic properties of a topology for which a distributed universally optimal strategy exists with $d.5$ rounds of message passing in a general $K$-user interference channel for $d > 1$ is a problem of great importance, and is still open.

## APPENDIX A
## UNIVERSALLY OPTIMAL STRATEGIES WITH 1.5 ROUNDS IN THREE USER TOPOLOGIES

In a three-user interference channel, there are at-most six cross links, existence or non-existence of which gives rise to $2^6 = 64$ cases. Some of the cases are topologically equivalent and hence that will reduce the total number of possibilities considered in this paper to 16 that are shown in Figure 1. Written below each figure is the number of topologies that are equivalent to that topology.

We now consider all these topologies one by one. We note that topologies (a), (b), (c), (d) and (p) satisfy the condition of universal optimality in the statement of the theorem and thus universal optimal strategy exist for only these topologies. We will now prove for the remaining topologies that there does not exist a universally optimal strategy.

**(e)**: Consider the strategy of the second user. It does not know any other direct link. If $n_{11} = n_{33} = 0$ and the second user did not transmit at rate $n_{22}$, its strategy will not be universally optimal. Thus, the second user sends at rate $n_{22}$ which is the only hope that a universally optimal strategy can be obtained. Now, consider that the channel gains in actual turn out to be all unity. Then, the optimal strategy for the first and the third user will be to turn off since the second user is transmitting at unit rate. Thus, the sum rate of 1 is achieved. However, the sum-capacity is 2 when the first and the third transmitter send at unit rate while the second user remains silent. Thus, there does not exist an universally optimal strategy.

**(f)**: Consider the strategy of the third user. It does not know any other direct link. Hence if $n_{22} = 0$ and the third user did not transmit at rate $n_{33}$, its strategy will not be universally optimal. Hence, the third user sends at rate $n_{33}$ which is the only hope that a universally optimal strategy can be obtained. Now, consider that the channel gains in actual turn out to be all unity. Then, the optimal strategy for the second user will be to turn off since $R_2 + R_3 \leq 1$. Further, the first user do not know $n_{33}$. So, it has to assume that the second user sends at $n_{22}$. If it does not, then there exist a case when $n_{33} = 0$ at which the second user will send at $n_{22}$ and the strategy of the first user will not be optimal. Thus in the above case of all links unity, the first transmitter will need to remain silent since it thinks that the second user sends at a unit rate and $R_1 + R_2 \leq 1$. However, the sum rate of 1 is achieved but the sum-capacity is 2 and thus there does not exist any universally optimal strategy.

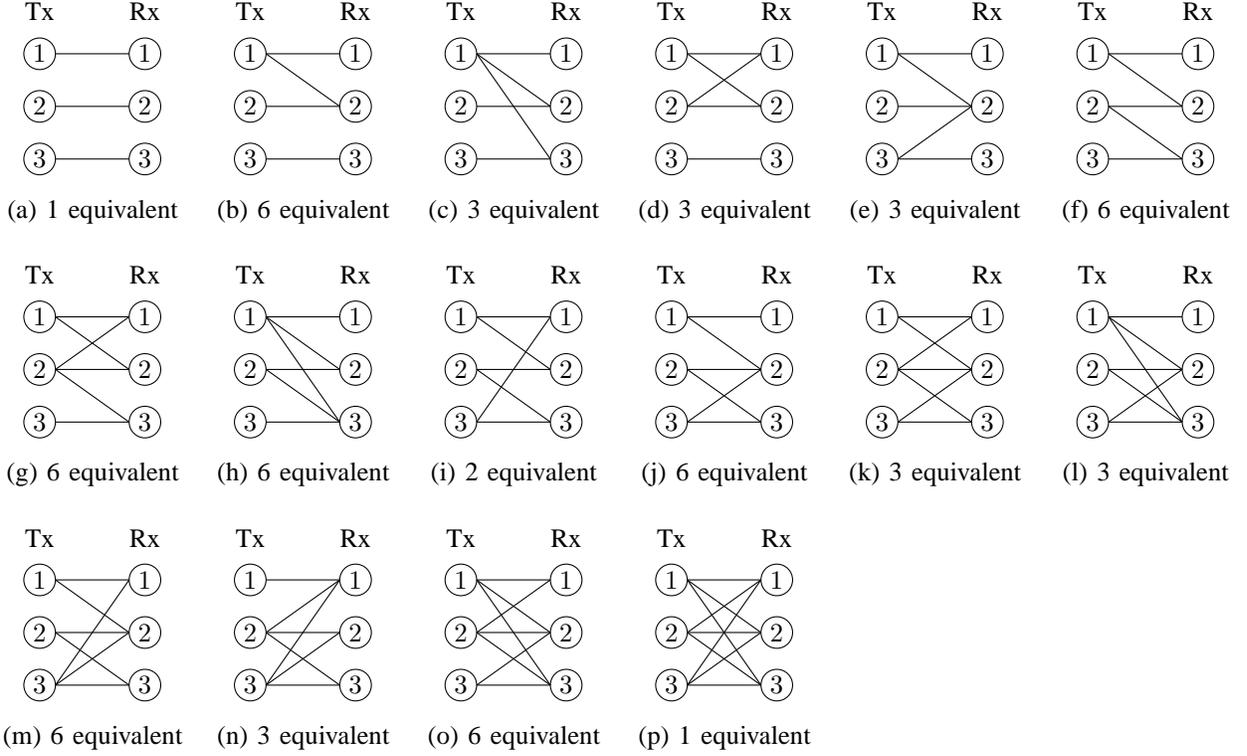

Fig. 1. The possible topologies in a three-user interference channel.

**(g)**: Consider the strategy of the third user. It does not know any other direct link. Hence if $n_{22} = 0$ and the third user did not transmit at rate $n_{33}$, its strategy will not be universally optimal. Hence, the third user sends at rate $n_{33}$ which is the only hope that a universally optimal strategy can be obtained. The first user does not know $n_{33}$. Now consider $n_{11} = n_{22} = 3$ and $n_{12} = n_{21} = 2$. In this case, the optimal sum rate $R_1 + R_2$ is outer-bounded by 4 and can only be achieved by the first and the second user transmitting at the rate of 2 each. Thus, from the point of view of the first user, if it sends at any rate different from 2, then the strategy is not universally optimal when $n_{33} = 0$. Hence, it transmits at rate 2. However, if the other channels turn out as $n_{23} = n_{33} = 3$. Then, the second user will have to keep silent as $R_3 = 3$ and $R_2 + R_3 \leq 3$. Thus, the sum rate of 5 is achieved. However, the fact that the second transmitter is silent is not known to the first transmitter and that is why it cannot increase its rate to 3 which it could do with global knowledge and thus there does not exist any universally optimal strategy.

**(h)**: Consider the strategy of the third user. It does not know any other direct link. Hence if $n_{22} = n_{11} = 0$ and the third user did not transmit at rate $n_{33}$, its strategy will not be universally optimal. Hence, the third user sends at rate $n_{33}$ which is the only hope that a universally optimal strategy can be obtained. Let the configuration of channels is $n_{11} = n_{22} = n_{33} = n_{13} = 6$, $n_{23} = 2$ and $n_{12} = 1$. The third user sends at $R_3 = 6$. For the second user, it does not know $n_{11}$. Also, $R_2 + R_3 \leq 10$ and thus if it sends at rate less than 4, the sum rate will not be optimal when $n_{11} = 0$. Thus, $R_2 = 4$. Further since $R_1 + R_3 \leq 6$, the first transmitter remains silent. Hence, the sum rate of 10 is achieved. However, $R_1 = 6$, $R_2 = 5$ and $R_3 = 0$ can be achieved which gives a higher sum rate which contradicts the universal optimality.

**(i)**: Consider the topology having all the direct links of weight 3 while the cross (or interfering) links of weight 2. Since it is symmetric, we consider only the first user. Since it does not know $n_{33}$, it may happen that $n_{33} = 0$. However if $n_{33} = 0$, the first transmitter knows that the second transmitter does not know $n_{11}$ and thus only possible universal optimal strategy for the second user to send at $n_{22}$. Since $R_1 + R_2 \leq 4$, it sends at a rate of 1. Similarly, the second and the third user send at unit rate. The sum rate of 3 is achieved. However, with global information, rate pair of $(3, 0, 1)$ can be achieving thus proving that there does not exist any universally optimal strategy for this topology.

**(j)**: Consider $n_{11} = n_{12} = n_{22} = n_{33} = 3$ and $n_{23} = n_{32} = 2$. For the third receiver, if it sends at any rate different from 2, it will not be optimal in case $n_{11} = 0$. From the point of view of the first user, if $n_{33} = 0$, the only possible universal strategy at the second receiver is to send at $n_{22}$ and thus the first transmitter should remain silent. Since, the second receiver does not know $n_{11}$, the optimal strategy for it is to use $R_2 = 2$; otherwise the sum rate is not optimal when $n_{11} = 0$. Thus, the sum rate of 4 is achieved. However, with global information the rate pair of $(3, 0, 3)$ can be achieved thus proving the claim.

**(k)**: Suppose that there exist a universally optimal strategy. When all the link gains $= 3$, $R_1 + R_2 \leq 3$ and $R_2 + R_3 \leq 3$. Thus, the optimal rate pair from sum rate is $(3, 0, 3)$. Any other rate pair in the region will give lower achievable rate. Hence, the first and the third user have to send at full rate if they see a two-user channel with all channel gains 3 in the universally optimal strategy. Now, consider the channel gains $n_{11} = n_{22} = n_{33} = n_{23} = n_{32} = 3$ and $n_{12} = n_{21} = 2$. Thus, the third transmitter sends at rate 3 as shown before. The first transmitter sends at rate 2; otherwise the rate will not be optimal when $n_{33} = 0$. Since $R_2 + R_3 \leq 3$, the second transmitter remains silent. Thus, the sum rate of 5 is achieved which is less than the sum rate of the pair $(3, 0, 3)$ that can be achieved will full information.

**(l)**: Consider $n_{11} = n_{22} = n_{33} = n_{13} = 6$, $n_{23} = n_{32} = 4$ and $n_{12} = 1$. Since, the second and the third user do not know $n_{11}$; only rate pair of $(4, 4)$ is optimal when $n_{11} = 0$. Thus, the second and the third user transmits at a rate of 4 each. The first transmitter knows the whole topology and since $R_1 + R_3 \leq 6$, the first transmitter sends at a maximum rate of 2. Thus, maximum sum rate with this strategy will be 10. However, the rate pair of $(5, 6, 0)$ can be achieved with full topology and thus there do not exist a universally optimal strategy.

**(m)**: Consider $n_{11} = n_{22} = n_{33} = n_{12} = 6$, $n_{23} = n_{32} = 4$ and $n_{31} = 1$. The first user sees this as a S-channel and therefore the only optimal choice for the first user is to send at a rate of 0 since the second user will be transmitting at full rate when $n_{33} = 0$. The second transmitter sees this as a two-user fully connected network and hence for $n_{11} = 0$, the only rate point that maximizes the sum rate is $R_2 = 4$. Since the third user knows the whole topology and $R_2 + R_3 \leq 8$, it can send at a maximum rate of 4. Thus, the maximum sum rate achieved is 8. However, the rate pair of $(5, 0, 6)$ can be achieved with full information thus proving the claim.

**(n)**: Consider $n_{11} = n_{22} = n_{33} = 4$, $n_{21} = n_{31} = 2$ and $n_{23} = n_{32} = 1$. The first user do not know $n_{22}$ and $n_{33}$ and thus it sends at rate $n_{11} = 4$; otherwise it won't be universally optimal when $n_{22} = n_{33} = 0$. The second and the third user know the topology. Since $R_1 + R_2 \leq 6$ and $R_1 + R_3 \leq 6$, the maximum sum rate of 8 can be achieved. However, the rate pair of $(3, 3, 3)$ can be achieved will full information thus proving the claim.

**(o)**: Consider $n_{11} = n_{22} = n_{33} = n_{12} = n_{21} = 6$, $n_{23} = n_{32} = 4$ and $n_{13} = 1$. Since the third transmitter does not know $n_{11}$ and only $R_3 = 4$ is optimal when $n_{11} = 0$, the third transmitter uses $R_3 = 4$. The first and the second transmitter knows the whole topology but $R_1 + R_2 \leq 6$. Thus, the maximum sum rate of 10 can be achieved. However with full information, the rate pair of $(5, 0, 6)$ can be achieved thus proving the claim.